\documentclass[%
reprint,
superscriptaddress,
pra,
]{revtex4-2}

\usepackage{graphicx} % Required for inserting images
\usepackage{amsmath} 
\usepackage[percent]{overpic}
\usepackage{xcolor}
\usepackage{braket}
\usepackage[colorlinks=true, linkcolor=red, anchorcolor=black, citecolor=blue,
urlcolor = blue]{hyperref}
\usepackage{ulem}

\begin{document}

\title{Sensitivity Comparison of Rydberg Atom-Based Radio-Frequency Electric Field Detection: Ionization Current Versus Optical Readout}

\author{Dangka~Shylla}
\affiliation{Department of Physics, University of Colorado, Boulder, Colorado 80302, USA}
\affiliation{ National Institute of Standards and Technology, Boulder, Colorado 80305, USA}
\author{Rajavardhan~Talashila}
\affiliation{Department of Electrical Engineering, University of Colorado, Boulder, Colorado 80309, USA}
\affiliation{ National Institute of Standards and Technology, Boulder, Colorado 80305, USA}
\author{Alexandra~Artusio-Glimpse}
\affiliation{ National Institute of Standards and Technology, Boulder, Colorado 80305, USA}
\author{Adil~Meraki}
\affiliation{Department of Physics, University of Colorado, Boulder, Colorado 80302, USA}
\affiliation{ National Institute of Standards and Technology, Boulder, Colorado 80305, USA}
\author{Dixith~Manchaiah}
\affiliation{Department of Physics, University of Colorado, Boulder, Colorado 80302, USA}
\affiliation{ National Institute of Standards and Technology, Boulder, Colorado 80305, USA}
\author{Noah~Schlossberger}
\affiliation{ National Institute of Standards and Technology, Boulder, Colorado 80305, USA}
\author{Samuel~Berweger}
\affiliation{ National Institute of Standards and Technology, Boulder, Colorado 80305, USA}
\author{Matthew~T.~Simons}
\affiliation{ National Institute of Standards and Technology, Boulder, Colorado 80305, USA}
\author{Christopher~L.~Holloway}
\affiliation{ National Institute of Standards and Technology, Boulder, Colorado 80305, USA}
\author{Nikunjkumar~Prajapati}
\affiliation{ National Institute of Standards and Technology, Boulder, Colorado 80305, USA}
 
\date{\today}

\begin{abstract}
We investigate a technique for detecting radio-frequency (RF) electric fields in a Cesium (Cs) vapor cell at room temperature by collecting charge from ionized Rydberg atoms and compare its performance with the established method of electromagnetically induced transparency (EIT). By applying a known RF field, we measure the response from both the electrical (ionization current-based) and optical (EIT-based) readouts. The ionization current-based method yields a sensitivity of 22~$\mu$Vm$^{-1}$Hz$^{-1/2}$, while the EIT-based method achieves 3.7~$\mu$Vm$^{-1}$Hz$^{-1/2}$. The sensitivity of the ionization current-based method is limited by thermal noise arising from a 2.2~k$\Omega$ resistance between the collection electrodes, attributed to a thin Cs film on the inner surfaces of the vapor cell. Controlling or eliminating the Cs layer can significantly improve the sensitivity of this ionization approach.

\end{abstract}

\maketitle

\section{Introduction}\label{sec:intro}

Detecting radio-frequency (RF) fields is important for applications like communications, radar, and precision sensing. Rydberg atom-based sensors have emerged as a promising platform for electric field metrology, offering absolute, SI-traceable measurements with high sensitivity across a broad frequency range in compact, room-temperature systems~\cite{Schlossberger2024Nature}. Various detection methods exist in the Rydberg sensor space. The most common detection method is the electromagnetically induced transparency (EIT), a nonlinear two-photon process where the transmission of an optical probe beam is modified via a coupling laser resonant with a Rydberg state. This process pumps atoms into the Rydberg state and allows for measurement of the Rydberg state energy. Any modification of this energy by an external field is determined by measuring the change in transmission of the probe laser~\cite{NikRepump21,Jing2020,Holloway22SRR,Cai2022,Bohaichuk2022,You22}. Other methods that utilize the two-photon process exists where the information carrier is the fluorescence emitted by the atoms rather than the transmitted signal~\cite{NikFluo24,Teale22}. In atomic beam lines and cold atom systems, selective field ionization has been utilized to read out a current to probe the energy and populations of several Rydberg states~\cite{Noah25Temp}. 

However, each of these methods may have potential limitations. EIT relies on the measurement of the probe laser transmission where the EIT contrast is small compared to the total transmission. This can potentially limit the sensitivity of such a device. Fluorescence measurements are limited in bandwidth by the lifetime of the state of interest. For receiver applications, this is stifling. Selective field ionization requires very large fields provided by plates and results in broad spectra with limited information for sensing.

\begin{figure}[h]
    \centering
        \begin{overpic}[width=1.0\columnwidth]{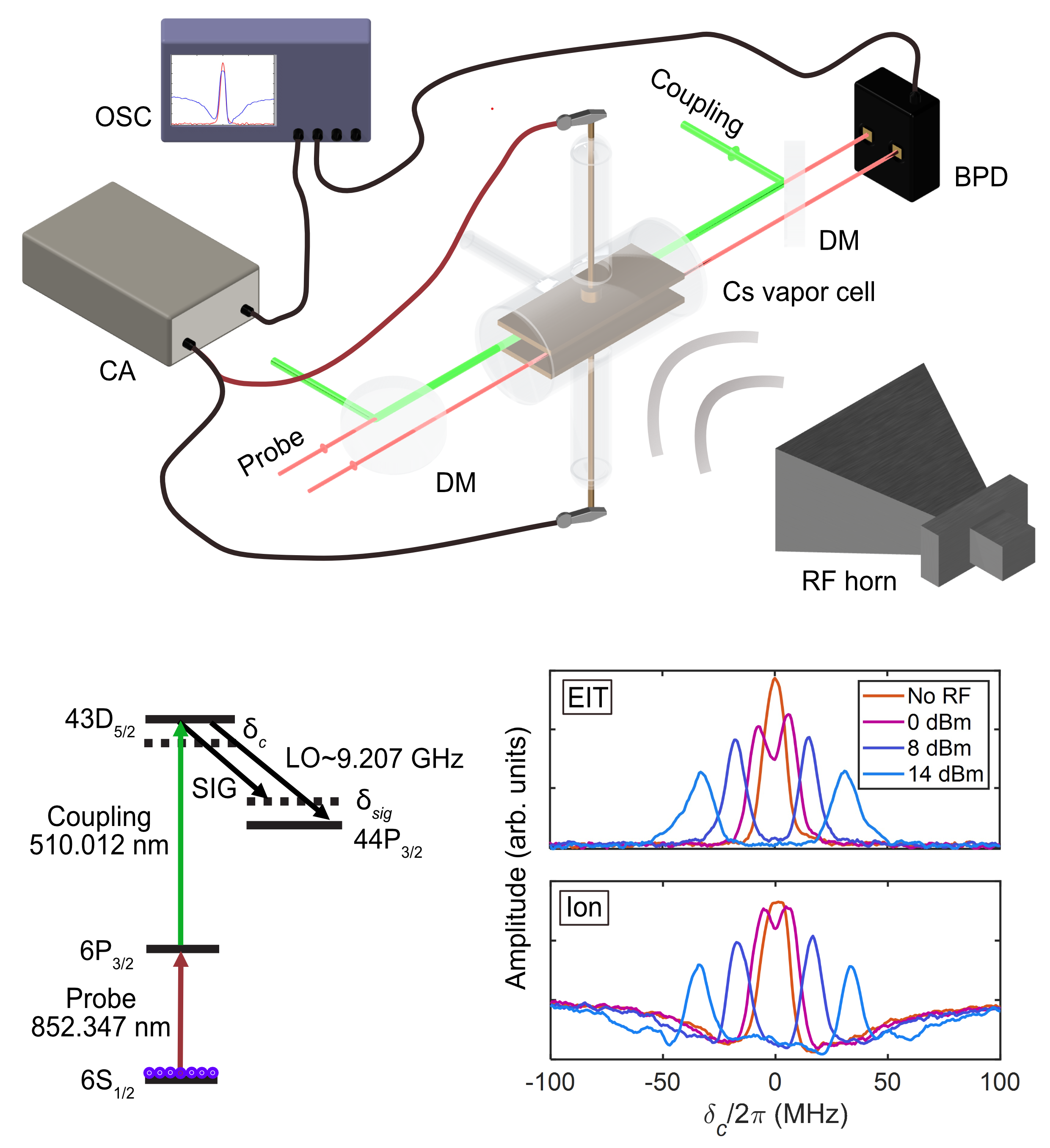}
        \put(1,100){\color{black}\textbf{(a)}}
        \put(1,45){\color{black}\textbf{(b)}}
        \put(45,45){\color{black}\textbf{(c)}}
    \end{overpic}
      \caption{(a) Schematic of the experimental setup. The probe and coupling laser beams counter-propagate through a Cs vapor cell positioned between parallel electrode plates. The LO and signal RF fields are radiated from a horn antenna and incident perpendicular to the laser beam paths. All fields are vertically polarized. DM: Dichroic mirror, BPD: Balanced photodetector, CA: Current amplifier and OSC: Oscilloscope. (b) Energy level diagram of $^{133}$Cs being probe by the experimental setup. The 852~nm probe laser is driving the $\text{6S}_{1/2}\rightarrow\text{6P}_{3/2}$ transition. The coupling laser at 510~nm drives the $\text{6P}_{3/2}\rightarrow\text{43D}_{5/2}$ Rydberg transition. The LO RF field resonantly drives the $\text{43D}_{5/2}\rightarrow\text{44P}_{3/2}$ Rydberg transition along with a weak RF signal detuned by $\delta_{sig}=10~\text{kHz}$ relative to the $\text{44P}_{3/2}$ state. (c) Example spectra showing EIT signal (top), obtained from the balanced photodetector receiving the transmitted probe beam, and the ionization signal (bottom), obtained from the current amplifier. Both signals are recorded simultaneously as a function of coupling laser detuning, $\delta_c$, for various injected signal RF powers to the cable feeding the horn antenna. The probe and coupling Rabi frequencies are fixed at $\Omega_p/2\pi = 3$~MHz and $\Omega_c/2\pi = 2.8$~MHz, respectively.}
    \label{fig:setup}
\end{figure}

In this work, we demonstrate a new method of readout relying on Rydberg state ionization via laser induced collisions. When atoms are excited to the Rydberg state, they are more susceptible to ionization via collisional effects. We can observe this ionization by utilizing a vapor cell embedded with stainless steel plates, shown in Figure~\ref{fig:setup}(a). As we scan the coupling laser across the Rydberg state resonance, we can observe the standard EIT from the two-photon resonance. At the same time, we also observe the ionization signal by measuring the current across the plates using a current amplifier, shown in Figure~\ref{fig:setup}(c). 
We emphasize the importance of the new readout method for the plated vapor cell. Previous studies have used vapor cells with internal electrodes to enable precision DC~\cite{Han2025} and low-frequency field sensing~\cite{HolV22,Xiao2025,Arumugam2025,Ma2022}, broadband detection via Stark tuning in the microwave regime~\cite{Ouyang2023}, and strong-field VHF measurements using Floquet analysis~\cite{Paradis2019}.

We characterize the RF electric field detection in a thermal cesium (Cs) vapor cell using an ionization current readout method. We evaluate the field sensitivity using a calibrated RF source and investigate its noise-limited performance. To benchmark the performance of this approach, we also perform measurements using the established EIT-based optical readout method under the same conditions. The results highlight the strengths and limitations of ionization-based detection and offer insight into its potential as a complementary technique to optical methods in Rydberg atom-based RF sensing.

%add ~5 ref fromAVS
\section{Methods}\label{sec:exp}
The experimental setup is shown in Figure~\ref{fig:setup}(a), where the probe and coupling laser beams counter-propagate through the Cesium (Cs) vapor cell positioned at the center between the parallel electrode plates. The Gaussian $1/e^2$ beam waist of the probe and coupling beams are fixed at $\sim$1~mm each. The local oscillator (LO) and signal RF fields, both vertically polarized along the same direction as the laser beams are radiated from a horn antenna and incident perpendicular to the laser beam propagation directions. The two RF fields are sourced from different signal generators that are synchronized with a 10~MHz reference clock and combined using a power combiner before being injected to the RF horn. The probe and coupling lasers are frequency-stabilized using ultra-low expansion (ULE) cavity lock.
%add linewidth, ule cavity locking

The experiment is carried out using a room temperature Cs vapor cell of length 5~cm and diameter of 2.5~cm that contains two parallel electrodes of stainless steel material embedded inside, also shown in Figure~\ref{fig:setup}(a). These electrodes, measuring 4 x 2~cm and separated by 5~mm are used to collect charges that originate from ionized Rydberg atoms through blackbody radiation-induced ionization, collisional ionization, photoionization, and field ionization \cite{galRyd94}. The collected charges produce an electrical current at the electrodes, which is fed into a low-noise current amplifier for measurement. 

The energy level scheme of $^{133}$Cs considered is shown in Figure~\ref{fig:setup}(b). The 852~nm probe laser is resonant with the $\mathrm{6S}_{1/2}(\text{F}=4) \rightarrow \mathrm{6P}_{3/2}(\text{F}'=5)$ transition, while the 510~nm coupling laser excites the $\mathrm{6P}_{3/2}(\text{F}'=5) \rightarrow \mathrm{43D}_{5/2}$ Rydberg transition, establishing a ladder-type EIT configuration. The two RF fields are configured to form a Rydberg atom mixer. The LO is set to $\sim$9.207~GHz, resonant to the $\mathrm{43D}_{5/2} \rightarrow \mathrm{44P}_{3/2}$ Rydberg transition. The weak signal field is detuned by $\delta_{sig} = 10~\text{kHz}$ relative to the $\mathrm{44P}_{3/2}$ Rydberg state. The interaction produces a beat note at the intermediate frequency or the frequency difference between the two RF fields i.e., $10$~kHz which is used for the RF electric field sensitivity measurements for both the optical EIT-based and ionization current-based methods. 

The transmitted 852~nm probe laser is detected using a balanced photodetector with a bandwidth of 10~MHz. The balanced detection helps suppress common-mode noise and enable a clean EIT signal readout. Meanwhile, the resulting ionization current is amplified by a low noise current amplifier with a bandwidth of 20~kHz. The beat note signal from both the optical and ionization channels are recorded using a spectrum analyzer with a frequency range of 9~kHz-7.1~GHz in the same vapor cell. The spectrum analyser is configured with a 10~Hz resolution bandwidth (RBW) and centered at the 10~kHz beat frequency generated when both the RF fields are switched on. 

The dipole matrix elements for the $\text{6S}_{1/2}\rightarrow\text{6P}_{3/2}$, $\text{6P}_{3/2}\rightarrow\text{43D}_{5/2}$, and $\text{43D}_{5/2}\rightarrow\text{44P}_{3/2}$ transitions are 2.0214$(ea_0)$, 0.0123$(ea_0)$, and 1195.7422$(ea_0)$, respectively~\cite{SIBALIC2017}, where $e$ is the elementary charge and $a_0$ is the Bohr radius. The typical linewidth of the first excited state, $6\text{P}_{3/2}$ is $\Gamma_2/2\pi= 5.2~\text{MHz}$~\cite{SteckCs98}, and those of the Rydberg states, $43\text{D}_{5/2}$ and $44\text{P}_{3/2}$ are $\Gamma_3/2\pi=5.1~\text{kHz}$ and $\Gamma_4/2\pi= 2.2~\text{kHz}$, respectively~\cite{Bet09}.

\section{Results}\label{sec:res}

\begin{figure}%[ht]
    \centering
    % Subfigure (a)
    \begin{overpic}[width=01\columnwidth]{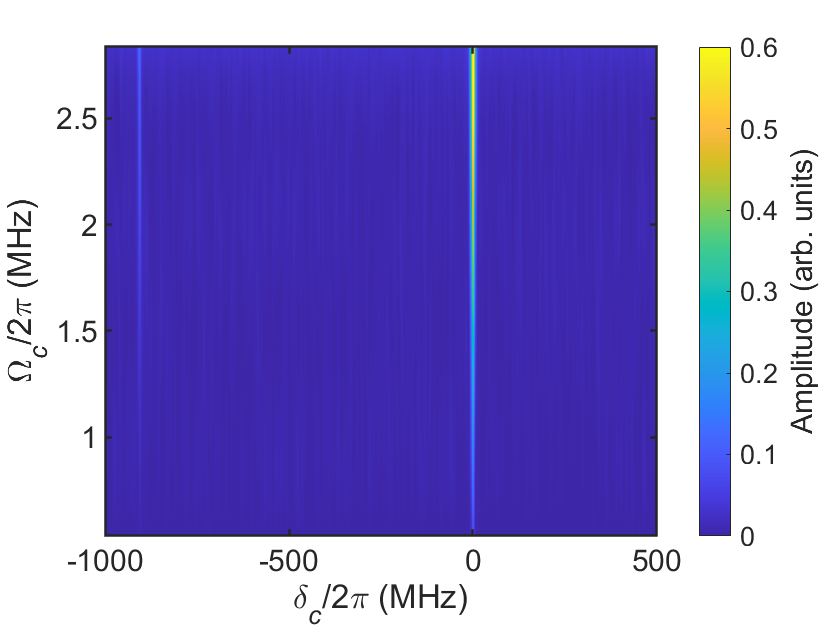}
        \put(1,72){\color{black}\textbf{(a)}}
    \end{overpic}
    \vspace{0em}  % vertical spacing between images
    % Subfigure (b)
    \begin{overpic}[width=01\columnwidth]{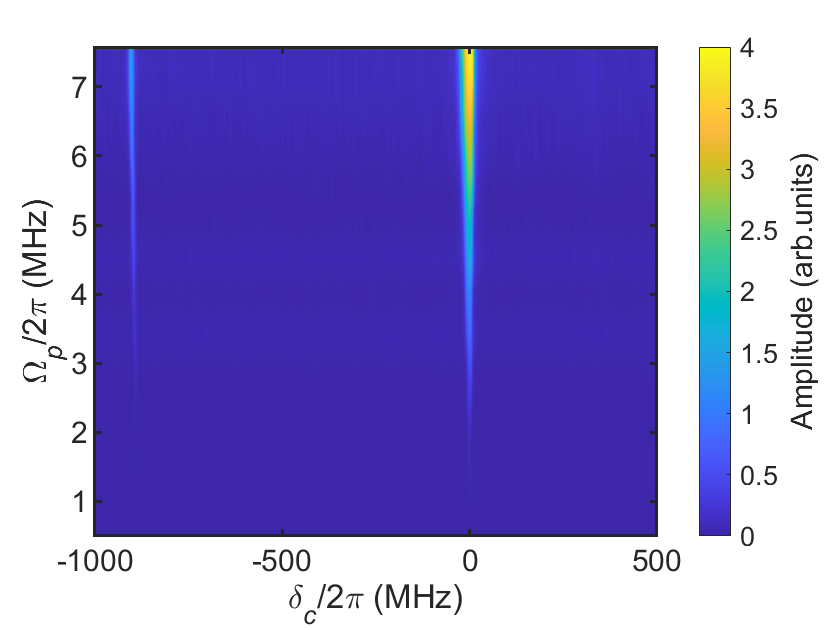}
        \put(1,72){\color{black}\textbf{(b)}}
    \end{overpic}
    \caption{
    EIT spectra versus coupling laser detuning, $\delta_c$, for (a) varying coupling Rabi frequencies, $\Omega_c$, at a fixed probe Rabi frequency of $\Omega_p/2\pi = 3$~MHz, and (b) varying probe Rabi frequencies, $\Omega_p$, at a fixed coupling Rabi frequency of $\Omega_c/2\pi = 2.8$~MHz. %The inset shows a zoomed-in view of the spectra at low probe Rabi frequencies.
    }
    \label{fig:EIT_Spectras}
\end{figure}

\begin{figure}%[ht]
    \centering
    % Subfigure (a)
    \begin{overpic}[width=1\columnwidth]{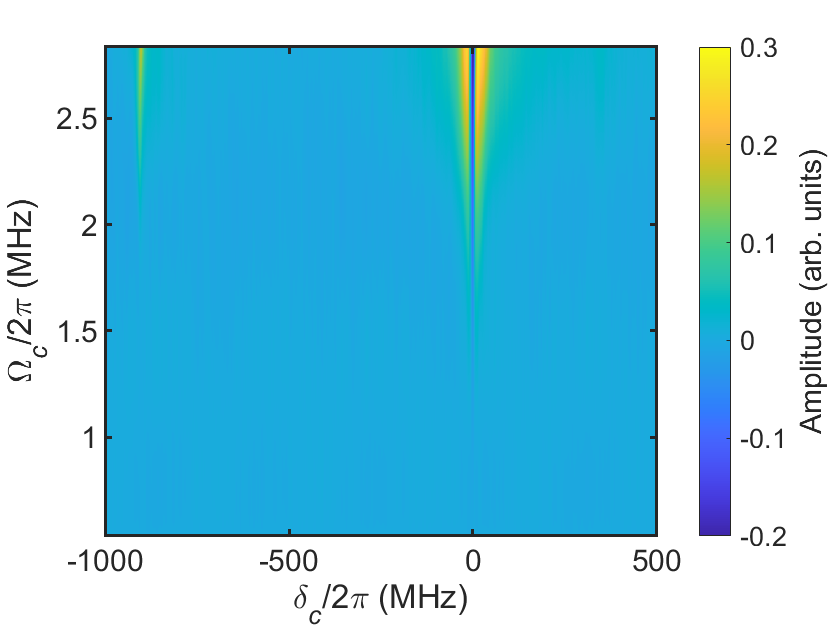}
        \put(1,72){\color{black}\textbf{(a)}}
    \end{overpic}
    \vspace{0em}  % vertical spacing between images
    % Subfigure (b)
    \begin{overpic}[width=1\columnwidth]{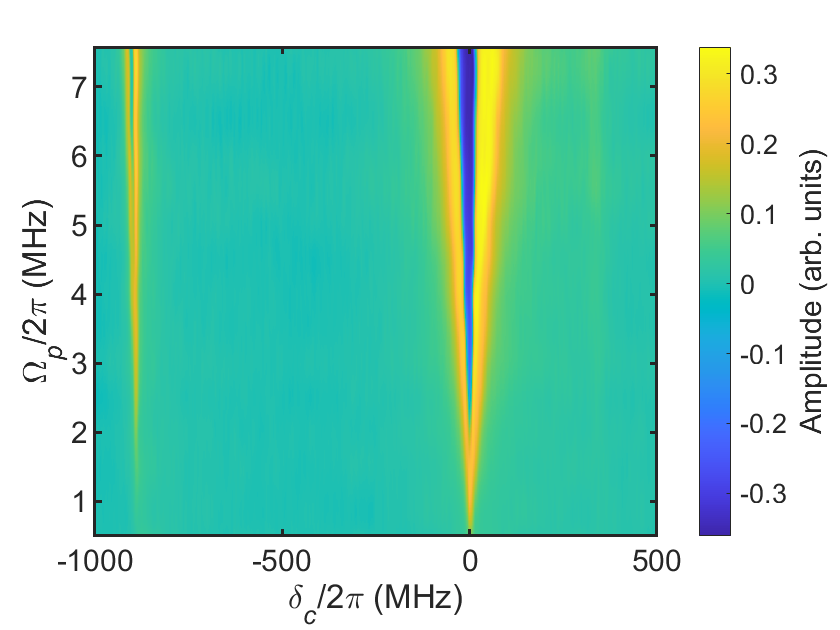}
        \put(1,72){\color{black}\textbf{(b)}}
    \end{overpic}
    \caption{Ionization spectra versus coupling laser detuning, $\delta_c$, for (a) varying coupling Rabi frequencies, $\Omega_c$, at a fixed probe Rabi frequency of $\Omega_p/2\pi = 3$~MHz, and (b) varying probe Rabi frequencies, $\Omega_p$, at a fixed coupling Rabi frequency of $\Omega_c/2\pi = 2.8$~MHz. %The inset highlights the transition region where the ionization signal flips from positive to negative with increasing probe Rabi frequency due to a change in which plate dominates the received signal.
    }
    \label{fig:Ion_Spectras}
\end{figure}

\begin{figure}%[ht]
    \centering
    % Subfigure (a)
    \begin{overpic}[width=1\columnwidth]{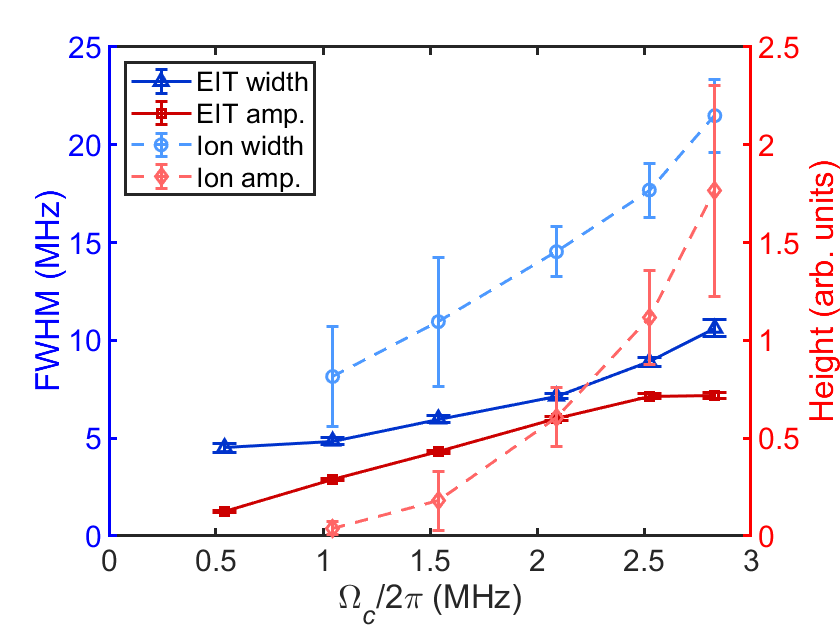}
        \put(1,72){\color{black}\textbf{(a)}}
    \end{overpic}
    \vspace{0em}  % vertical spacing between images
    % Subfigure (b)
    \begin{overpic}[width=1\columnwidth]{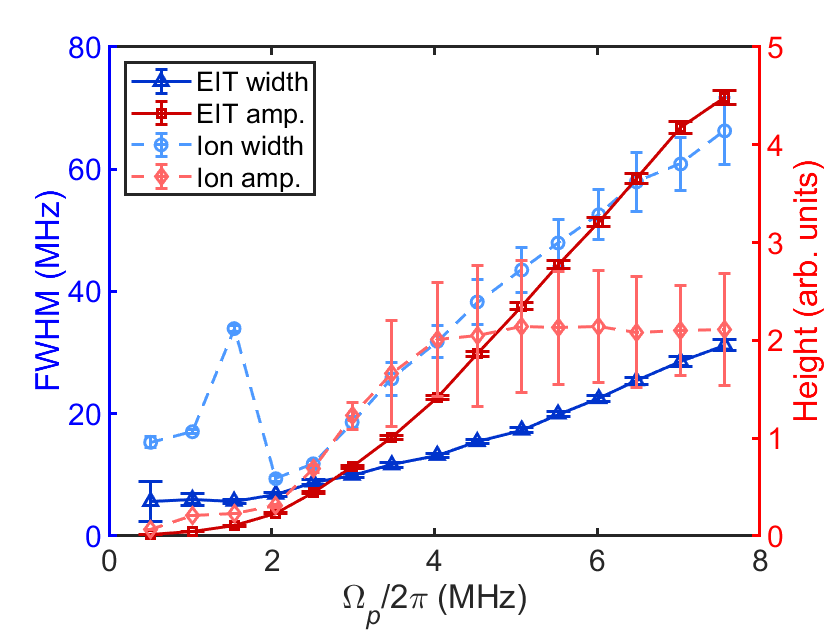}
        \put(1,72){\color{black}\textbf{(b)}}
    \end{overpic}
    \caption{Full width at half maximum (FWHM) and amplitude versus (a) coupling Rabi frequency at a fixed probe Rabi frequency of $\Omega_p/2\pi = 3$~MHz, and (b) probe Rabi frequency at a fixed coupling Rabi frequency of $\Omega_c/2\pi = 2.8$~MHz for both EIT and ionization spectra. Error bars show fitting uncertainties (1$\sigma$).}
    \label{fig:LWnAmp}
\end{figure}

We begin by plotting the EIT spectra as a function of coupling laser detuning for various coupling and probe Rabi frequencies using colormap plots as shown in Figure~\ref{fig:EIT_Spectras}~{(a) and (b)}, with the RF fields turned off. The spectra show two peaks corresponding to Rydberg states $43\text{D}_{3/2}$ (left) and $43\text{D}_{5/2}$ (right). Similarly, we plot the corresponding ionization spectra under the same conditions and are shown in Figure~\ref{fig:Ion_Spectras}~{(a) and (b)}. Their linewidth and peak height are extracted from Lorentzian fits to the EIT spectra and from double Lorentzian dip fits to the ionization spectra. The extracted values are plotted in Figure~\ref{fig:LWnAmp}(a) and (b) as a function of probe and coupling Rabi frequencies, respectively. 

As the coupling Rabi frequency increases, both the EIT and ionization signals increase in width and height in a straight-forward manner. When the probe Rabi frequency is increased, the EIT peak height increases and broadens as a result of power broadening. In contrast, the ionization signal shows a polarity reversal in the peaks, followed by an increase in width and the dip height which eventually saturates. The extracted linewidth shows an initially increase with the probe Rabi frequency but after the polarity flip, the linewidth decreases before increasing again. As the probe Rabi frequency increases, the fit quality for the ionization spectra deteriorates, as indicated by an increased uncertainty (1$\sigma$) in the fit.

This behavior is likely due to the absence of an externally applied bias field, which would otherwise guide the charges to a specific electrode. Instead, we see a variation in which electrode the charges travel to. At low probe Rabi frequencies, fewer Rydberg atoms are excited, resulting in a weaker ionization signal. As the probe Rabi frequency increases, the number of ionized Rydberg atoms grows, and the imbalance in charge collection becomes more pronounced, leading to the observed feature in the ionization signal.%, see the inset of Figure~\ref{fig:Ion_Spectras}~(b) for a close view of this transition.

%Without such a field, the collected signal varies depending on which electrode the charges reach.
%highlight the curve in the plot for the best sensitivities
%add F=2,3 and so on in figure 1b

\begin{figure}%[ht]
    \centering
    \begin{overpic}[width=1.05\columnwidth]{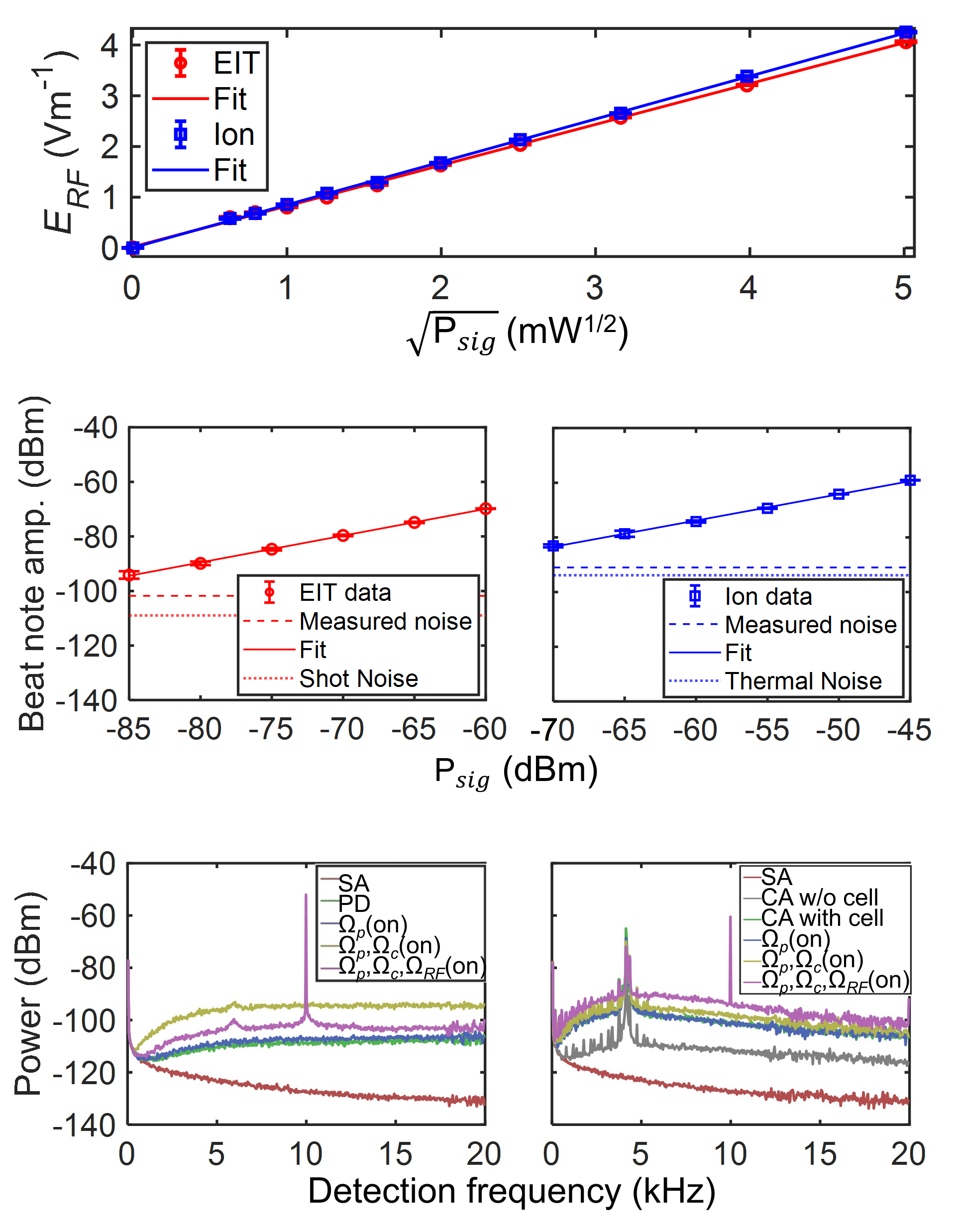}
        \put(1,100){\color{black}\textbf{(a)}}
        \put(1,68){\color{black}\textbf{(b)}}
        \put(42,68){\color{black}\textbf{(c)}}
        \put(1,32){\color{black}\textbf{(d)}}
        \put(42,32){\color{black}\textbf{(e)}}
    \end{overpic}
    \vspace{0em}  % vertical spacing between images
    \caption{(a) Measured RF electric field amplitude, $\text{E}_{RF}$, versus the square root of input signal RF power, $\sqrt{\text{P}_{sig}}$ in mW$^{1/2}$, used to extract the calibration factor via linear fitting. Solid red circles and green squares data points correspond to EIT and ionization signals, respectively, with corresponding red and green lines indicating their linear fits. Beat note amplitude as a function of input signal RF power, $\text{P}_{sig}$ in dBm for (b) EIT, and (c) ionization. The solid line indicates the linear fit, the dashed line marks the measured noise floor, and the dotted line represents the calculated noise level. Noise spectra for (d) EIT, and (e) ionization: the solid red curve shows the spectrum analyzer (SA) noise, the green curve shows the balanced photodetector noise for EIT and current amplifier noise for ionization (with the cell connected). The blue curve shows probe laser noise alone, the yellow curve corresponds to probe laser noise with the coupling laser and the purple curve shows probe laser noise with both coupling and RF fields. The gray curve (ionization only) shows current amplifier noise with the input open.}
    \label{fig:Cali_noise}
\end{figure}

Next, we calibrate the signal RF field strength from the Autler-Townes (AT) spectra as a function of coupling laser detuning for various signal RF powers, as shown in Figure~\ref{fig:setup}~(c) for both EIT and ionization readouts. This calibration is necessary because the actual RF field experienced by the atoms inside the vapor cell differs from the nominal output power due to losses in the cable between the signal generator and horn antenna, propagation of the field from the antenna to the cell, and reflections caused by the glass cell. By using the AT splitting, we directly probe the field as experienced by the atoms. This is done by calibrating the coupling laser frequency scan using the known 903.17~MHz \cite{SIBALIC2017} separation between the nearby $43\text{D}_{3/2}$ and $43\text{D}_{5/2}$ Rydberg states, as shown in Figures~\ref{fig:EIT_Spectras} and \ref{fig:Ion_Spectras}. The extracted AT splitting, $\delta_{{AT}}$, is then used to calculate the RF electric field amplitude using:
\begin{equation}
E_{RF} = \frac{h \delta_{{AT}}}{\mu_{34}}~,
\end{equation}
where $\mu_{34}=1195.742(ea_0)$ is the transition dipole moment of the $\mathrm{44D}_{5/2} \rightarrow \mathrm{43P}_{3/2}$ Rydberg transition and $h$ is the Planck's constant. We obtain the calibration factor, $C_{fac}$, for both the EIT and ionization readouts from linear fits of $E_{RF}$ versus $\sqrt{P}$ in $\mathrm{mW}^{1/2}$ as shown in Figure~\ref{fig:Cali_noise}~(a), which relates the applied RF power to the RF field strength inside the cell~\cite{And14}. For EIT and ionization readouts, the extracted calibration factors are 0.78 and 0.83~Vm$^{-1}\text{mW}^{-1/2}$ respectively. The small difference in the calibration factors may be attributed to minor day-to-day alignment variations between the EIT and ionization sensitivity measurements, though the two remain in close agreement.  %This enables accurate and quantitative sensitivity measurements in absolute physical units.

After obtaining the calibration factor, we measure the sensitivity of the system for both the EIT and ionization readouts using the atom-based mixer approach, where a beat note at $10$~kHz is generated and detected by a spectrum analyzer. The LO RF power is swept, and the corresponding 10~kHz beat note amplitude is recorded. The optimal LO power is determined as the power at which the beat note amplitude reaches its maximum. To measure the noise floor, the signal RF is turned off and the LO RF remains on and the spectral power near 10~kHz is recorded across five repetitions and averaged. Next, the signal RF is turned on and swept over a range of powers. For each power level, the beat note amplitude at 10~kHz is extracted and averaged over five traces. These amplitudes are plotted against the input signal RF power and fitted with a linear function as shown in Figure \ref{fig:Cali_noise}(b) and (c). The minimum detectable RF field corresponds to the smallest signal level at which the beat note can be distinguished from background noise. This threshold is defined as the input signal RF power required to achieve a signal-to-noise (SNR) of one, i.e., the point where the beatnote amplitude equals the average noise floor. This power is extracted from the linear fit and converted to field strength using the calibration factor, $C_{{fac}}$, constant obtained from the AT splitting measurements. The corresponding sensitivity is then calculated using:
\begin{equation}
{S} ={C}_{{fac}}\cdot\sqrt{\frac{10^{{{P}_{sig}^{min}}/{10}}}{\delta_{RBW}}}~,
\end{equation}
where ${P}_{sig}^{min}$ is the minimum detectable power in dBm defined by $\text{SNR}=1$, and $\delta_{RBW}=$ 10~Hz is the configured spectrum analyzer resolution bandwidth. The resulting sensitivity is expressed in units of Vm$^{-1}\text{Hz}^{{-}1/2}$, representing the minimum detectable RF field normalized to a 10~Hz detection bandwidth. 

The sensitivity of measurement techniques is ultimately limited by fundamental noise sources. Figure~\ref{fig:Cali_noise} (d) and (e) shows the measured noise spectra for the EIT and ionization channels, respectively, with contributions from various sources. For the optical (EIT-based) detection system, the dominant limitation is the photon shot noise, which produces root mean squared (RMS) voltage fluctuations at the photodetector output given by:
\begin{equation}
{V}_{rms}^{shot} =  {G}_{{PD}}\cdot \sqrt{2 e{I}_{{ph}} \delta_{RBW}}~,
\end{equation}
where ${G}_{PD} = 630 \times 10^3\,\mathrm{VA^{-1}}$ is the transimpedance gain of the photodetector, ${I}_{{ph}} = \eta {P} / {E}_{{ph}}$ is the average photocurrent for incident optical probe power, ${P}$. The probe power measured at the photodetector is ${P} = 1~\mu\mathrm{W}$. ${E}_{{ph}} = hc/\lambda_p$ is the photon energy, $c$ is the speed of light, $\lambda_p = 852.347~\mathrm{nm}$ is the probe laser wavelength. $\eta = 0.73$ is the photodetector quantum efficiency. The corresponding noise power at the spectrum analyzer input is:
\begin{equation}
{P}_{{noise}}^{shot} = 10 \log_{10} \left( \frac{({V}_{rms}^{shot})^2}{{Z}_{{SA}} \cdot 10^{-3}} \right) = -108.9~\mathrm{dBm}~,
\end{equation}
where ${Z}_{{SA}} = 50~\Omega$ is the spectrum analyzer input impedance. The measured noise floor of $-103~\mathrm{dBm}$ indicates additional low-frequency ($1/f$) noise beyond the shot-noise limit, as shown in Figure~\ref{fig:Cali_noise}~(b).

For the electrical (ionization current) detection method, the primary limitation is thermal (Johnson–Nyquist) noise across the $2.2~\mathrm{k}\Omega$ resistive path between the collection plates. After amplification by a transimpedance amplifier with gain $\text{G}_{\text{CA}} = 0.5 \times 10^6\,\mathrm{VA^{-1}}$, the
RMS voltage noise at the output is:
\begin{equation}
{V}_{rms}^{thermal} =  {G}_{{CA}}\cdot \sqrt{\frac{4 k_B {T} \delta_{RBW}}{{R}}}~,
\end{equation}
where $k_B$ is the Boltzmann constant, ${T} = 300~\mathrm{K}$, and ${R} = 2.2~\mathrm{k}\Omega$. The resulting noise power at the spectrum analyzer is given by the same form:
\begin{equation}
{P}_{noise}^{thermal} = 10 \log_{10} \left( \frac{({V}_{rms}^{thermal})^2}{{Z}_{{SA}} \cdot 10^{-3}} \right) = -94.2~\mathrm{dBm}~.
\end{equation}
The calculated value agrees well with the measured noise level, confirming that thermal noise is the dominant limitation, as shown in Figure~\ref{fig:Cali_noise}~(b).

\begin{figure}%[ht]
    \centering
    % Subfigure (a)
    \begin{overpic}[width=1\columnwidth]{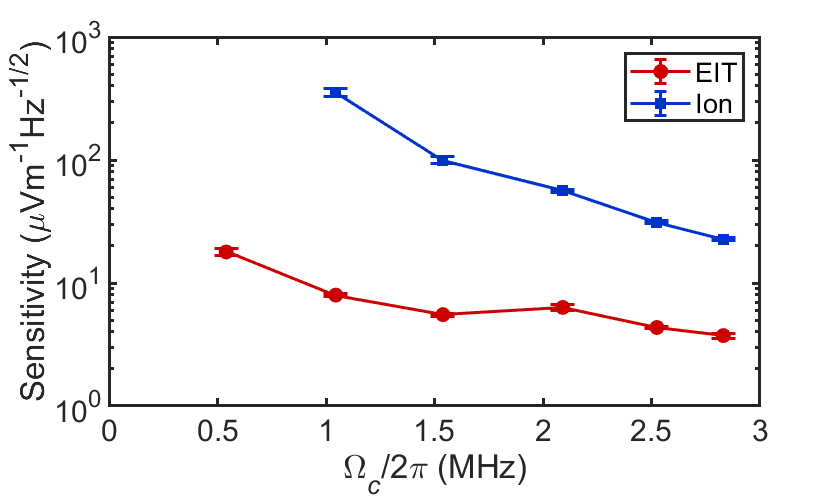}
        \put(1,60){\color{black}\textbf{(a)}}
    \end{overpic}
    \vspace{0em}  % vertical spacing between images
    % Subfigure (b)
    \begin{overpic}[width=1\columnwidth]{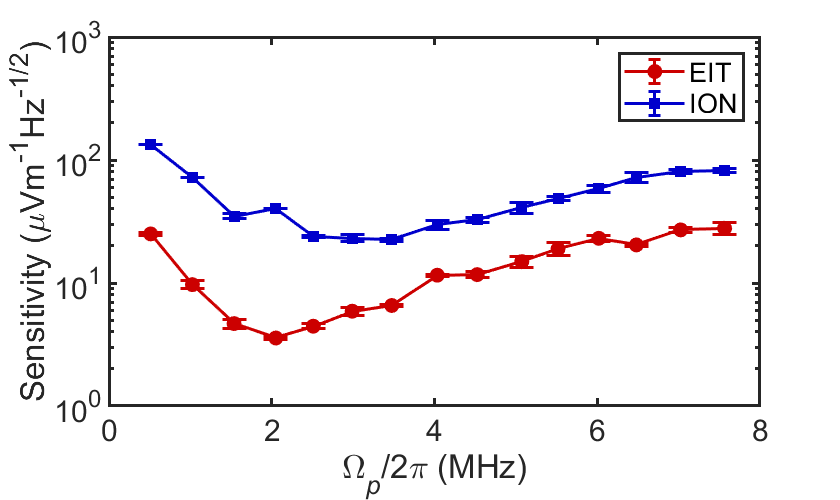}
        \put(1,60){\color{black}\textbf{(b)}}
    \end{overpic}
    \caption{Sensitivity in $\mu\text{Vm}^{-1}\text{Hz}^{-1/2}$ as a function of (a) coupling Rabi frequency, $\Omega_c$, with the probe Rabi frequency fixed at $\Omega_{p}/2\pi = 2.5$~MHz, and (b) probe Rabi frequency, $\Omega_p$, with the coupling Rabi frequency fixed at $\Omega_{c}/2\pi = 2.8$~MHz. Red circles report data points for EIT, and green squares report data points for ionization. Each data point represents the average of three independent measurements, with error bars indicating the standard deviation.}
    \label{fig:Sensitivities}
\end{figure}

Finally, we compare the sensitivities of both the optical (EIT-based) and electrical (ionization current) detection methods. Figures~\ref{fig:Sensitivities}~(a) and (b) shows the sensitivity, expressed in units of $\mu\text{V}\text{m}^{-1}\text{Hz}^{-1/2}$ as a function of the coupling and probe Rabi frequencies, respectively. As the coupling Rabi frequency increases, the sensitivity improves for both methods, consistent with enhanced coherence in the EIT process and increased ionization efficiency. In contrast, sensitivity as a function of the probe Rabi frequency shows an initial improvement with increasing power, followed by degradation due to power broadening effects. The optimal sensitivities achieved are $3.7~\mu\text{V}\text{m}^{-1}\text{Hz}^{-1/2}$ at $\Omega_p=2~\textrm{MHz},\Omega_c=2.8~\textrm{MHz}$ for the EIT-based method and $22~\mu\text{V}\text{m}^{-1} \text{Hz}^{-1/2}$ at $\Omega_p=3.5~\textrm{MHz},\Omega_c=2.8~\textrm{MHz}$ for the ionization current detection method. The lower sensitivity in the ionization-based approach is limited by thermal noise arising from a $ 2.2~\text{k}\Omega $ resistive path between the electrodes, attributed to a thin, conductive Cs film deposited on the inner surfaces of the vapor cell. This elevated noise floor is evident in Figure~\ref{fig:Cali_noise}~(e), where the current amplifier noise with no input (effectively infinite input resistance) is approximately $-110\,\mathrm{dBm}$, as shown by gray curve. In contrast, when the vapor cell, with its $2.2~\text{k}\Omega$ resistance, is connected to the amplifier input, the noise floor increases to approximately $-93~\mathrm{dBm}$, as shown by green curve, confirming the dominance of thermal noise in the ionization detection method.

\section{Discussion}\label{sec:Dis}
Efforts to optimize the electrical (ionization current-based) method for sensitivity measurement included heating the vapor cell to remove Cs atoms deposited on the glass walls and trapping them in the stem, especially near the regions where the glass connects to the electrodes. This was intended to increase the resistance between the plates and thereby reduce the thermal noise floor. However, this approach was ineffective, as the resistance consistently returned to $\approx2.2~\mathrm{k}\Omega$ after the vapor cell was left to cool~\cite{Asao32,Ive25}.
Additionally, a bias voltage was applied across the electrodes to externally guide charges to a specific electrode, but instead the ionization signal degraded due to the 2.2~k$\Omega$ resistance which limited the electric field between the plates, making the charge collection inefficient.
Future improvements could include using a vapor cell with a more controlled Cs quantity, as excessive Cs lowers the resistance between the plates.

\section{Conclusion}\label{sec:conc}
We have experimentally compared two Rybderg atom-based techniques for RF electric field detection in a thermal Cs vapor cell: ionization current and EIT readouts. Our measurements shows that
the EIT-based technique has better sensitivity, achieving  3.7~$\mu$Vm$^{-1}$Hz$^{-1/2}$, compared to 22~$\mu$Vm$^{-1}$Hz$^{-1/2}$ for the ionization current-based readout.The primary limitation of  the ionization current-based approach arises from thermal noise associated with a low-resistance path between the electrodes due to a conductive Cs film on the inner surfaces of the vapor cell. These results demonstrate that with proper control over cell conditions, ionization-based methods could offer a viable, compact alternative to optical readout in Rydberg atom-based field sensing systems.

\section*{Acknowledgments}
This research was developed with funding from  NIST under the NIST-on-a-Chip program. % A contribution of the U.S. government, this work is not subject to copyright in the United States. 

\subsection*{Conflict of Interest}
\vspace{-3mm}
The authors have no conflicts to disclose.

\vspace{-3mm}
\subsection*{Data Availability Statement}
\vspace{-3mm}
The data that support the findings of this study are openly available in NIST at \href{https://doi.org/10.18434/mds2-3951}{\textcolor{blue}{https://doi.org/10.18434/mds2-3951}}~\cite{Shylla2025Dataset}.

\bibliography{cites}

\end{document}